\documentclass[A4paper, 10pt, twocolumn]{article}
\usepackage{authblk}
\usepackage{amsmath,amsfonts,amssymb,amsthm,mathrsfs,empheq, nccmath}
\usepackage{subcaption}
\usepackage{hyperref}
\usepackage{url}
\usepackage{tikz}
\usepackage{xcolor}
\usepackage{multicol}
\usepackage{cuted}

\usepackage[margin=2cm]{geometry}

\title{Caputo Fractional Standard Map: Scaling Invariance Analyses}

\author[1]{Daniel Borin}

\affil[1]{Departamento de Física, Univ. Estadual Paulista - Unesp, Av.24A, 1515, Bela Vista, Rio Claro, CEP: 13506-900, SP, Brazil}

\date{}

\begin{document}
	
\twocolumn[
\begin{@twocolumnfalse}	
\centering Chaos, Solitons and Fractals \textbf{181} (2024) 114597 \\ \url{https://doi.org/10.1016/j.chaos.2024.114597}

\maketitle

\vspace{-1cm}
\begin{abstract}

\noindent In this paper, we investigate the scaling invariance of survival probability in the Caputo fractional standard map of the order $1<\alpha<2$ considered on a cylinder. We consider relatively large values of the nonlinearity parameter $K$ for which the map is chaotic. The survival probability has a short plateau followed by an exponential decay and is scaling invariant for all considered values of $\alpha$ and $K$.
\end{abstract}


\vspace{1cm}

  \end{@twocolumnfalse}
]

\section{Introduction}

Differential equations are a fundamental tool for studying dynamical systems, enabling the investigation of the evolution of observables over time and capturing the long-term behavior of these systems. These equations elucidate the rates of change in system variables and typically involve integer-order derivatives. A generalization of this theory has led to the rise of the field of fractional differential equations (FDEs) \cite{KST06}, extending the conventional framework of differential equations by incorporating fractional derivatives. This extension has countless applications in various scientific fields, such as quantum mechanics \cite{LXW23,LXW21,AS20, Laskin17, Lim06, Laskin02}, electrodynamics \cite{SG20, SG19, SP18, ML17}, economics \cite{Tarasov_Economia21, Tarasov_Economia20_1, Tarasov_Economia20_2, Tarasov_Economia19}, engineering \cite{CL19, YWHW15, WY14}, biology \cite{MBKJ23, JBAJ23}, and geology \cite{WBA20}.

\par Fractional derivatives come in various types, each with its unique characteristics and applications. The most common among them include Riemann-Liouville, Caputo, Grünwald-Letnikov, Hadamard type, etc. \cite{KST06}. As initial conditions are crucial in physical dynamics, adopting the Caputo derivative in mechanical systems is convenient. This choice is particularly useful since the initial conditions for specific fractional dynamical systems align with those for usual dynamical systems, allowing for a clearer physical interpretation of the system.

\par Sometimes, differential equations can be complicated. A way to simplify the problem is to reduce it to the study of simple discrete maps. This approach provides a simplified yet valuable perspective on the dynamics of the system, enhancing its suitability for simulations and computational studies. Maps, derived from equations of motion involving fractional derivatives, exhibit a noteworthy characteristic: long-term memory effects \cite{Tarasov09, TZ08}. In this context, the present state of the system is uniquely determined by all past states, adding an intriguing layer of complexity and historical dependence to the analysis. 

Memory is a significant property of human beings and is the subject of extensive biophysical and psychological research. The possibility of investigating the dynamics of memory in neurons and proteins \cite{protein} becomes attractive for the study of fractional maps.

\par An essential and insightful measure when exploring the statistical characteristics of transport phenomena is the concept of survival probability. This probability quantifies the likelihood of a diffusing particle remaining within a specific region at a given point in time. In the context of fully chaotic systems, the behavior of survival probability exhibits a distinct pattern over time, characterized by exponential decay \cite{Daniel1}. These properties have not been explored for Caputo fractional maps.

\par Motivated by the above descriptions, this paper aims to explore the dynamic characteristics of a Caputo Fractional Map obtained by replacing the integer derivative with the Caputo derivative in the equation of motion that originates the famous standard map (Chirikov-Taylor Map). More specifically, we vary control parameters and observe how they affect the survival probability, seeking invariance of scaling.

\par The layout of the study is as follows: The Caputo Fractional Standard Map is introduced in Sect. \ref{Sec:Deduction}. In section \ref{Sec:InvSca}, a study of survival probability is presented, varying all control parameters and demonstrating that this observable is scaling invariant concerning the same parameters. Finally, a discussion is provided in Section \ref{Sec:Discussion}.

\section{The Mapping} \label{Sec:Deduction}

The standard map (also known as Chirikov-Taylor Map) \cite{Chirikov_1, Chirikov_2}
\begin{equation}\label{1}
	\left\{\begin{array}{ll}
		p_{n+1}=p_n-K\sin(x_n) \\
		x_{n+1}=x_n +p_{n+1} \ \ (\text{mod }2\pi)
	\end{array} \right.
\end{equation}
is the solution of the equation
\begin{equation}\label{2}
	\ddot{x}=-K\sin(x) \sum_{n=0}^{\infty} \delta\left(\dfrac{t}{T}-n\right)	
\end{equation}
In this scenario, the perturbation involves a recurring sequence of delta-function-like pulses, or "kicks," with a periodicity of $T$ and an amplitude represented by $K$.

\par We can generalize this model by substituting the second derivative with a Caputo fractional derivative. The left-sided Caputo fractional derivative \cite{KST06,MR93} of order $\alpha>0$ is defined by

\begin{equation}\label{3}
		_0^C D_t^\alpha	f(t)  \coloneqq   \dfrac{1}{\Gamma(l-\alpha)} \int_0^t \dfrac{f^{(l)}(\tau)d\tau}{(t-\tau)^{\alpha-l+1}}.
\end{equation}

\noindent where $l \in \mathbb{N}$ such that $l-1<\alpha \leq l$. Note that when $\alpha=m\in \mathbb{N}$, the Caputo Fractional derivative of order $\alpha$ becomes an $m$th usual derivative, i.e., 
$$_0^C D_t^m	f(t)=\frac{d^m f(t)}{dt^m} \equiv f^{(m)}(t), \qquad \text{for }m \in \mathbb{N}. $$
In this way, the fractional Caputo generalization of \eqref{2} is given by
\begin{equation}\label{4}
	_0^C D_t^\alpha x(t)+K \sin(x(t)) \sum_{n=0}^{\infty} \delta\left(\dfrac{t}{T}-n\right)=0
\end{equation}

We can reduce  the Caputo fractional differential equations to a Caputo Generalization nonlinear Volterra integral equations of second kind \cite{KM05, KM04}, such that $x(t)$ is a solution of \eqref{4} if and only if it is a solution of the integral equation
\begin{equation}\label{5}
	\begin{array}{ll}
&x(t) \displaystyle =\sum_{k=0}^{l-1} \dfrac{x^{(k)}(0)}{k!}t^k \\
& \displaystyle +\dfrac{1}{\Gamma(\alpha)} \int_0^t (t-\tau)^{\alpha-1} \left[K \sin(x(\tau)) \sum_{j=0}^{\infty}\delta\left(\dfrac{\tau}{T}-j\right)\right] d\tau
\end{array}
\end{equation}
where $l-1<\alpha \leq l$. Consider $1<\alpha\leq2$, thus the Equation \ref{5} is equivalent to the discrete map equations \cite{Tarasov09, TZ08}

\begin{equation}\label{19}
\small\left\{ \begin{array}{ll}
		p_{n+1}=p_0-\dfrac{KT^{\alpha-1}}{\Gamma(\alpha-1)}\displaystyle \sum_{j=0}^{n} (n+1-j)^{\alpha-2} \sin(x_j) \\
		x_{n+1}=x_0+p_0(n+1)T-\dfrac{KT^{\alpha}}{\Gamma(\alpha)} \displaystyle\sum_{j=0}^{n}(n+1-j)^{\alpha-1} \sin(x_j) \\ (\text{mod } 2\pi)  
	\end{array} \right.
\end{equation}
for $1<\alpha\leq 2$. The Eq. \eqref{19} is named the Caputo Fractional Standard Map (CFSM). Note that if $\alpha = 2$, then Equation \eqref{19} gives the usual map \eqref{1}, making the CFSM a generalization of the standard map. The CFSM has no periodicity in $p$ and can be considered on a cylinder.

The CFSM may generate periodic sinks, attracting slow-diverging trajectories, attracting accelerator mode trajectories, chaotic attractors, a cascade of bifurcations, and an inverse cascade of bifurcations type of attracting trajectories have been reported in several studies \cite{Edelman11, Edelman13, ET13,  Edelman14, Edelman18, Edelman19, Edelman22}.

Self-similarity, distribution of exit times, and survival probability in the regular standard map were investigated in \cite{ZEN97}. However, the scattering properties of the CFSM map have not been explored yet. Thus, in the next section, we undertake this task and characterize the escape of orbits from the phase space of the CFSM.

\section{Scattering properties: scaling invariance of survival probability} \label{Sec:InvSca}

\hspace{0.5cm} The concept of scaling invariance is a well-known idea across various scientific and mathematical domains. It delves into the notion that certain phenomena can remain unchanged even when their parameters undergo rescaling or transformation. Essentially, if a system exhibits scaling invariance, its expected behaviors stay consistent regardless of scale. This characteristic is observed in diverse systems such as chaotic dynamics \cite{Daniel1}, human dynamics \cite{BHG06} and social networks \cite{OCL18}. 

\par The main goal of this section is the study of the survival probability of a trajectory remaining inside a specific region in the space phase, considering the CFSM Map proposed in the previous section. More specifically, our focus is on in investigating whether such behavior exhibits scale invariance when subjected to certain alterations of the mapping parameters.

\begin{figure*}[!ht]
	\centering
	\begin{subfigure}[ht]{0.48\textwidth}
		\centering
		\includegraphics[width=\textwidth]{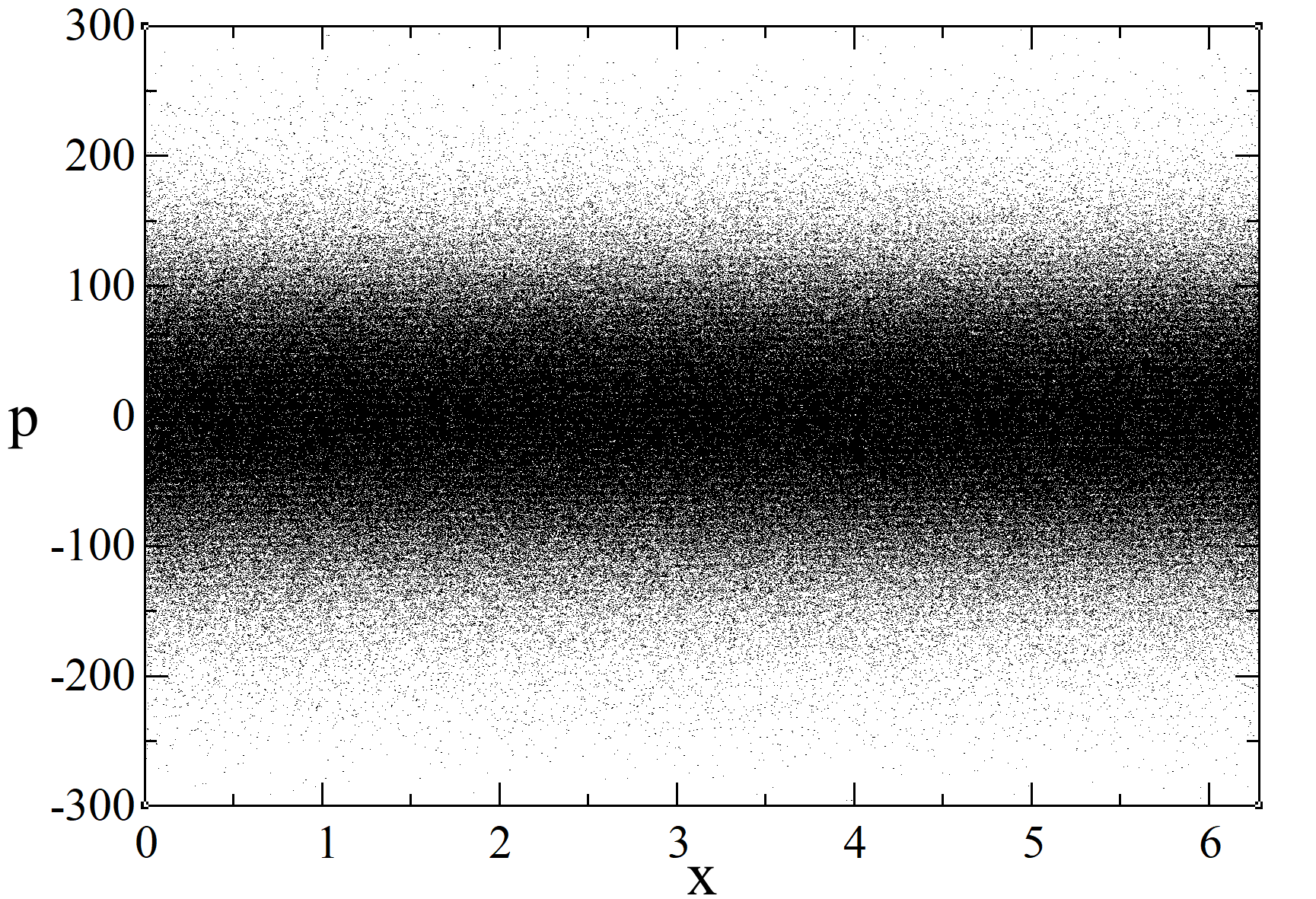}
		\caption{ }
		\label{Fig:SpaPha_K23,06a1,7}
	\end{subfigure}
	\hfill
	\begin{subfigure}[ht]{0.48\textwidth}
		\centering
		\vspace{0.23cm}
		\includegraphics[width=\textwidth]{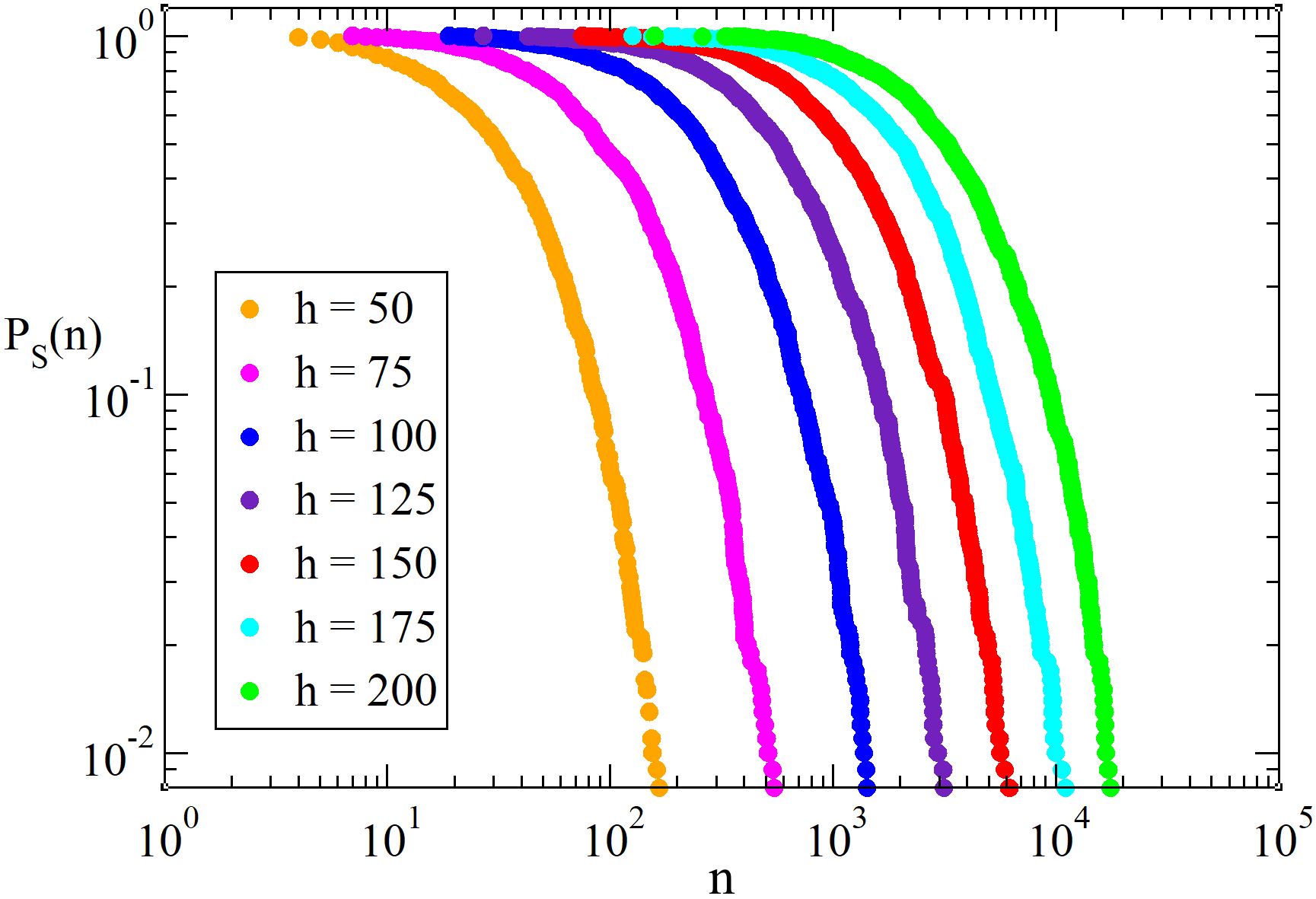}
		\caption{ }
		\label{Fig:ScalingInv1-a}
	\end{subfigure}
	\hfill
	\begin{subfigure}[ht]{0.48\textwidth}
		\centering
		\includegraphics[width=\textwidth]{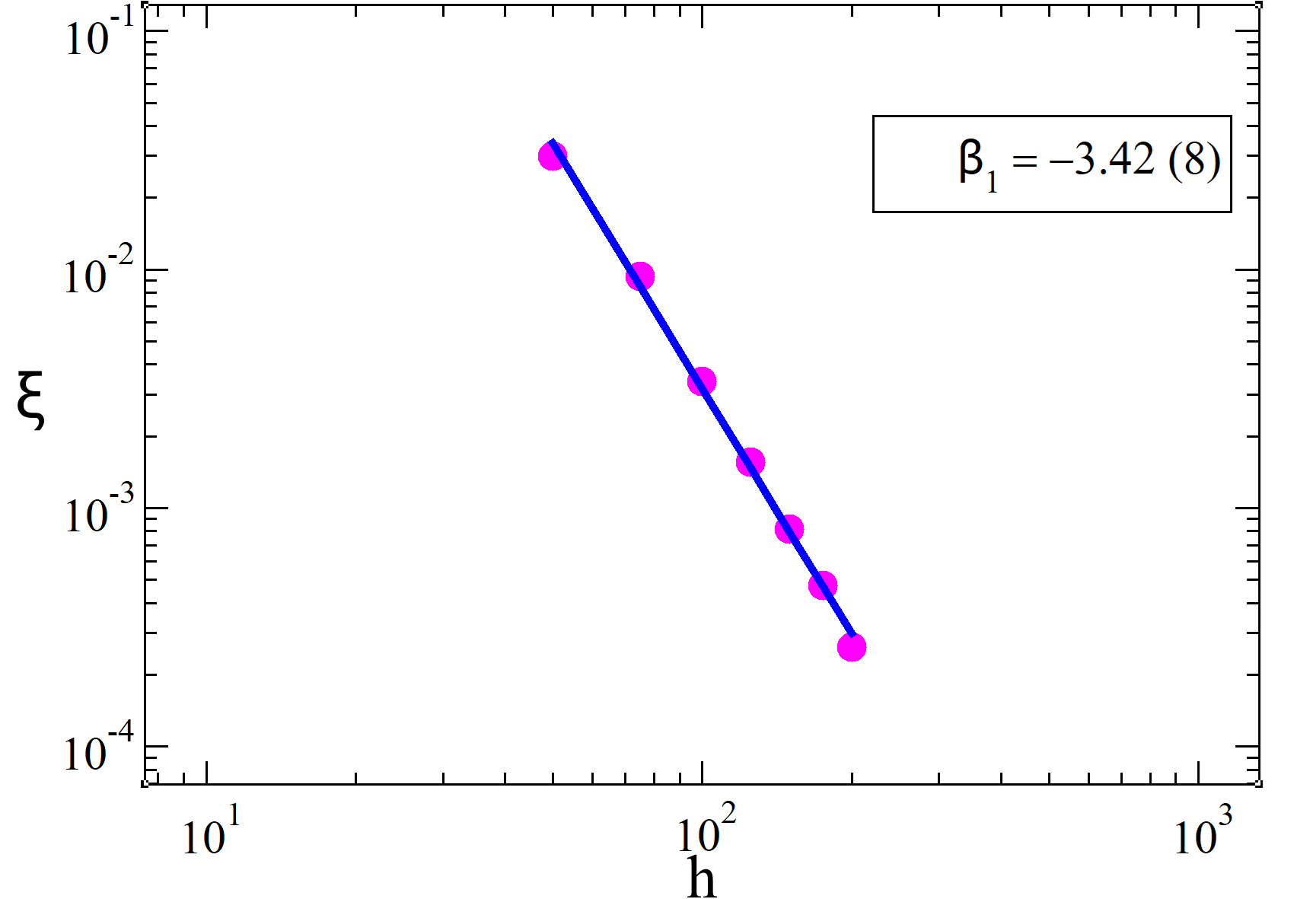} 
		\caption{ }
		\label{Fig:ScalingInv1-b}
	\end{subfigure}
	\hfill
	\begin{subfigure}[ht]{0.48\textwidth}
		\centering
		\includegraphics[width=\textwidth]{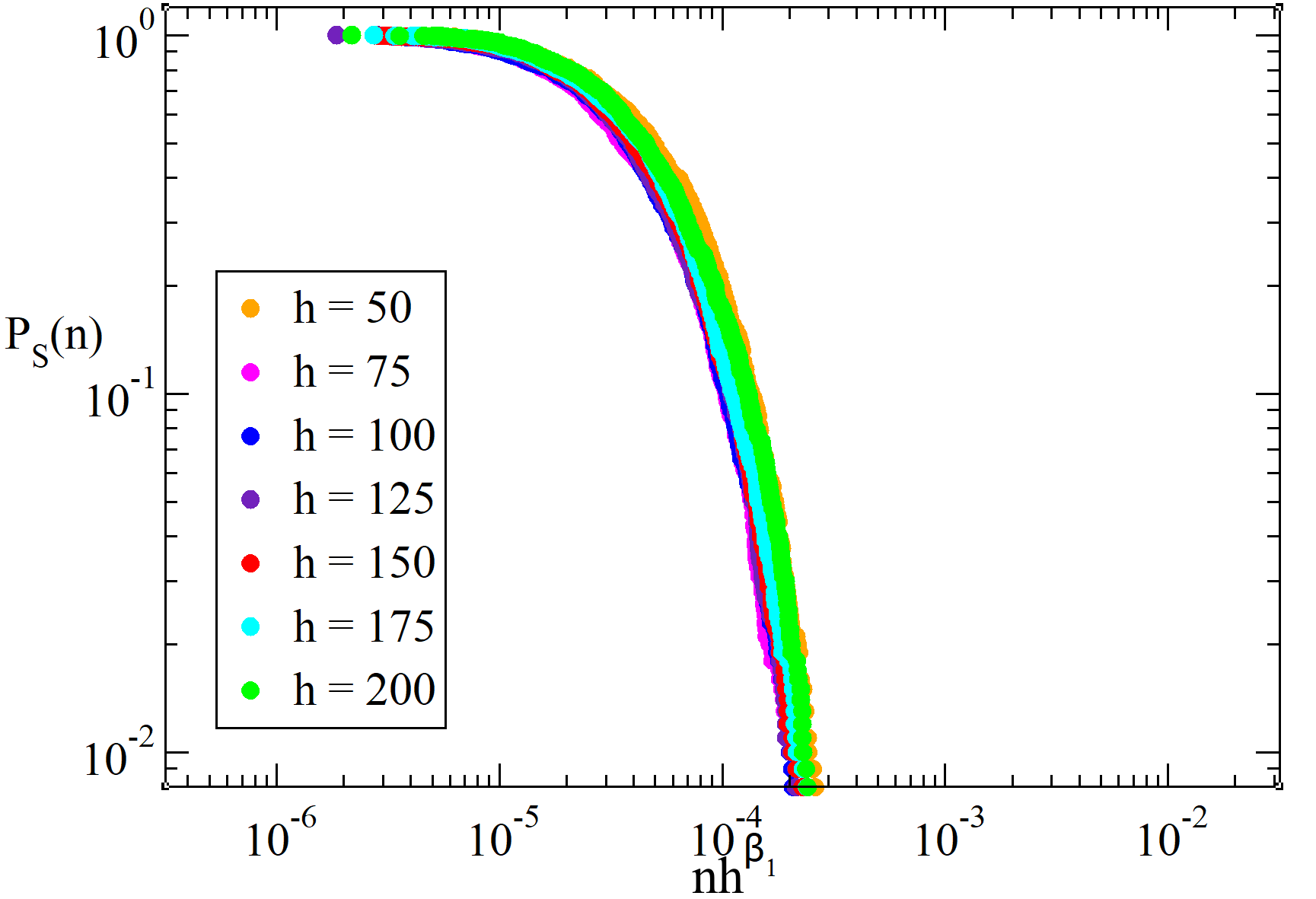}
		\caption{ }
		\label{Fig:ScalingInv1-c}
	\end{subfigure}
	\caption{(a) Space Phase $(x,p)$ for $K=23.0602$ and $\alpha=1.7$; (b) Plot of the survival probability curves $P_S (n)$ as functions of the number of collisions $n$, considering survival region height in the range $h\in[25,200]$ computed for $M=10^3$ orbits. The curves can be described by an exponential $P_S (n) \propto \exp(-\xi n)$; (c) The behavior of the critical exponent $\xi$ as a function of $h$. As can be seen, the relationship between height $h$ and critical exponent $\xi$ is given by a power law, such that $\xi  \propto h^{\beta_1}$, with an exponent $\beta_1=-3.42(8)$; (d) We display the overlapping of the $P_S(n)$ curves in a single universal curve, indicating the scaling invariance of the survival probability for the variations of survival region height $h$.}
\end{figure*}

\par To begin the analysis, it is essential to define the object of study: the measure of the probability of survival, denoted as $P_S(n)$ \cite{Bermudez}. This probability refers to the possibility of a given particle remaining inside of a determined region after $n$ interactions. Numerically, this measure is computed as \cite{Daniel1}
\begin{equation}
	P_{S} (n)=\dfrac{1}{M}\sum_{j=1}^n H_S (M,j), \qquad \text{for } n\in \mathbb{N} <N,
\end{equation}
where $H_S(M,j)$ denotes the histogram associated with $M$ distinct particles released in $j$ collisions. To calculate this observable, an ensemble of $M$ distinct orbits is considered, with initial conditions set as $x_0=0$ and random $p_0$ uniformly distributed in the interval $(0,2 \pi/100)$. Each orbit evolves according to CFSM \eqref{19} until the trajectory escapes a rectangular region characterized by $p \leq |h|$. Thus the number of interactions up to that instant is recorded, and then a new particle in the system is initialized. This a procedure is repeated several times until the entire ensemble of initial conditions is exhausted. Here, it is important to mention that in our approach, each particles has the opportunity to evolve within the system for up to $N = 10^6$ interactions if it has not escaped before.

\par With the incorporation of this new characterization of the dynamics, which is the survival region, three parameters, namely $K$, $\alpha$, and $h$, influence the study of scattering properties. The goal from now on is to examine the influence of these parameters on the survival probability and determine whether it exhibits the property of scale invariance.

\begin{figure*}[!ht]
	\centering
	\begin{subfigure}[ht]{0.48\textwidth}
		\centering
		\includegraphics[width=\textwidth]{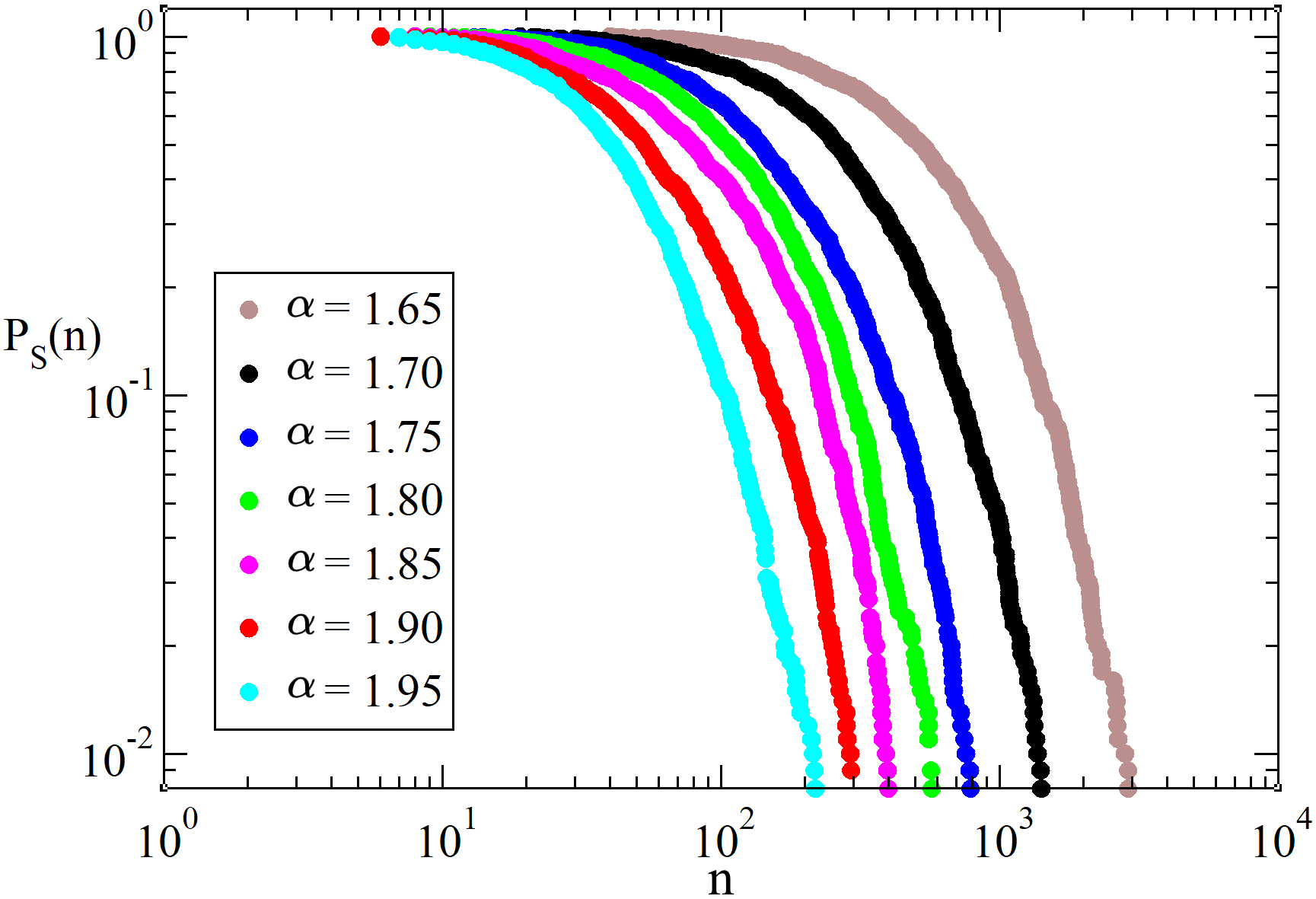}
		\caption{ }
		\label{Fig:ScalingInv2-a}
	\end{subfigure}
	\hfill
	\begin{subfigure}[ht]{0.48\textwidth}
		\centering
		\vspace{0.01cm}
		\includegraphics[width=\textwidth]{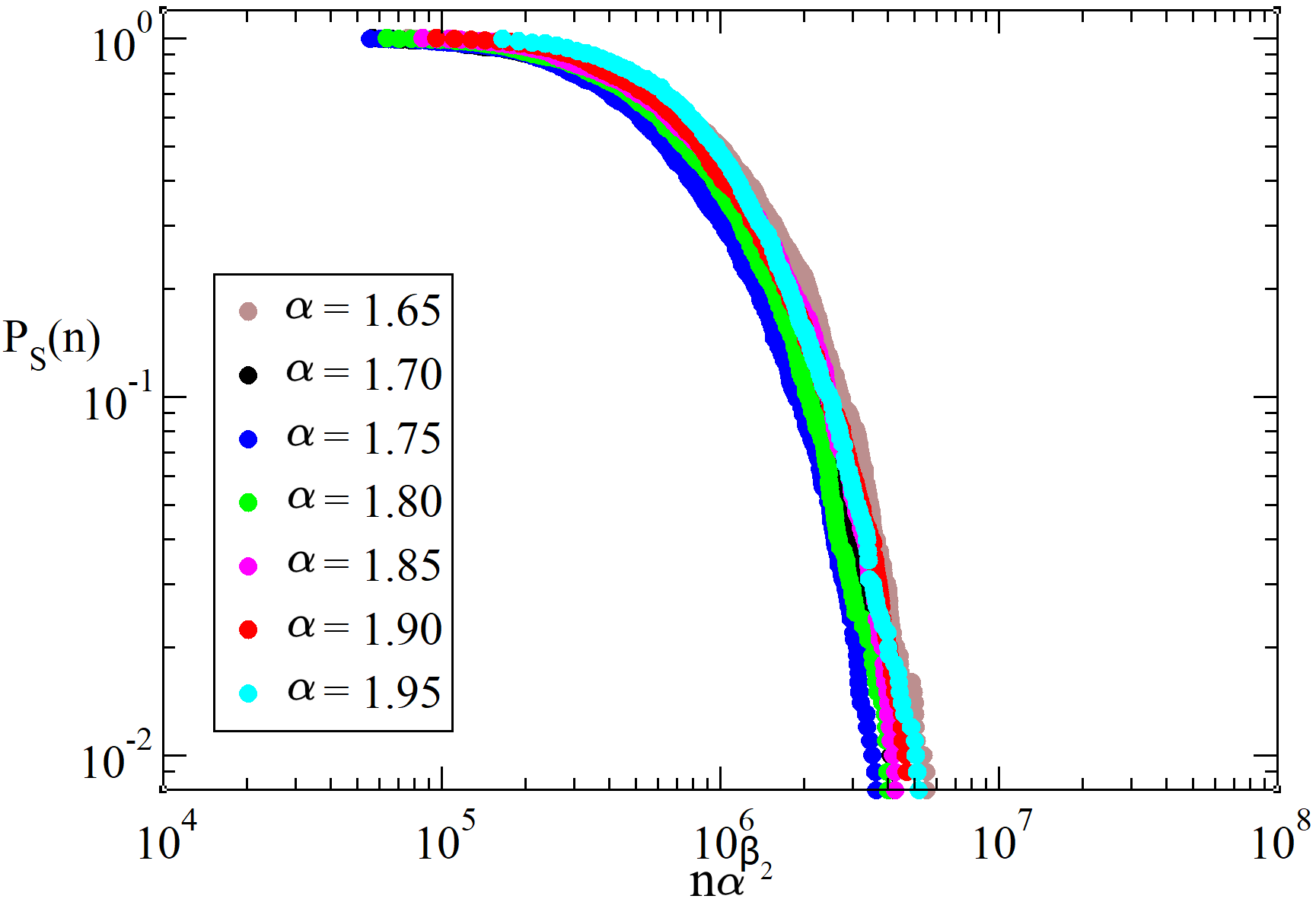}
		\caption{ }
		\label{Fig:ScalingInv2-c}
	\end{subfigure}
	\caption{(a) Plot of the survival probability curves $P_S (n)$ as functions of the number of collisions $n$, considering $\alpha$ in range the $\alpha\in[1.65,1.95]$, $K=23.0602$, $h=100$ and computed for $M=10^3$ orbits. The curves can be described by an exponential $P_S (n) \propto \exp(-\xi n)$; (b) We display the overlapping of the $P_S(n)$ curves in a single universal curve, indicating the scaling invariance of the survival probability concerning variations of $\alpha$.}
\end{figure*}

\par To avoid the cascade of bifurcation-type trajectories (CBTT) \cite{Edelman13b}, which exists only in fractional dynamical systems, larger values of $K$ will be considered, ensuring that the dynamics are essentially chaotic. The influence of these new types of attractors on the scattering properties will be investigated in future works.

\par Starting by keeping $K=23.0602$ and $\alpha=1.7$ constants, the phase space of \eqref{19} is illustrated in Fig. \ref{Fig:SpaPha_K23,06a1,7} and the survival probability $P_S(n)$ in the function of the number of interactions $n$ considering different sizes for the survival region height $h$, as displayed in Fig. \ref{Fig:ScalingInv1-a}. One can note that for medium and long values of interactions $P_S$ obeys an exponential decay as
\begin{equation}\label{key}
	P_S(n) \propto e^{-\xi n}.
\end{equation}
Furthermore, the behavior of $P_S$ depends on the value of $h$ assumed such that, from a logical point of view, the higher the value of $h$, the slower the decay of $P$ occurs, suggesting a relationship between $h$ and $\xi$ which is evident in Fig. \ref{Fig:ScalingInv1-b}, showing a power law function 
$$\xi \propto h^{\beta_1},$$
where $\beta_1=-3.42(8)$, $(\cdot)$ denotes the error of the previous digit. The knowledge of $\beta_1$ allows us to re-scale the horizontal axis as $n \rightarrow n h^{\beta_1}$ producing an overlap of the curves plotted in Fig. \ref{Fig:ScalingInv1-a} into a single and universal plot indicating that the exponential decay rate is scaling invariant with size of survival regions, as shown in Fig. \ref{Fig:ScalingInv1-c}.

\par
\begin{figure}[ht]
	\centering
		\includegraphics[width=0.48\textwidth]{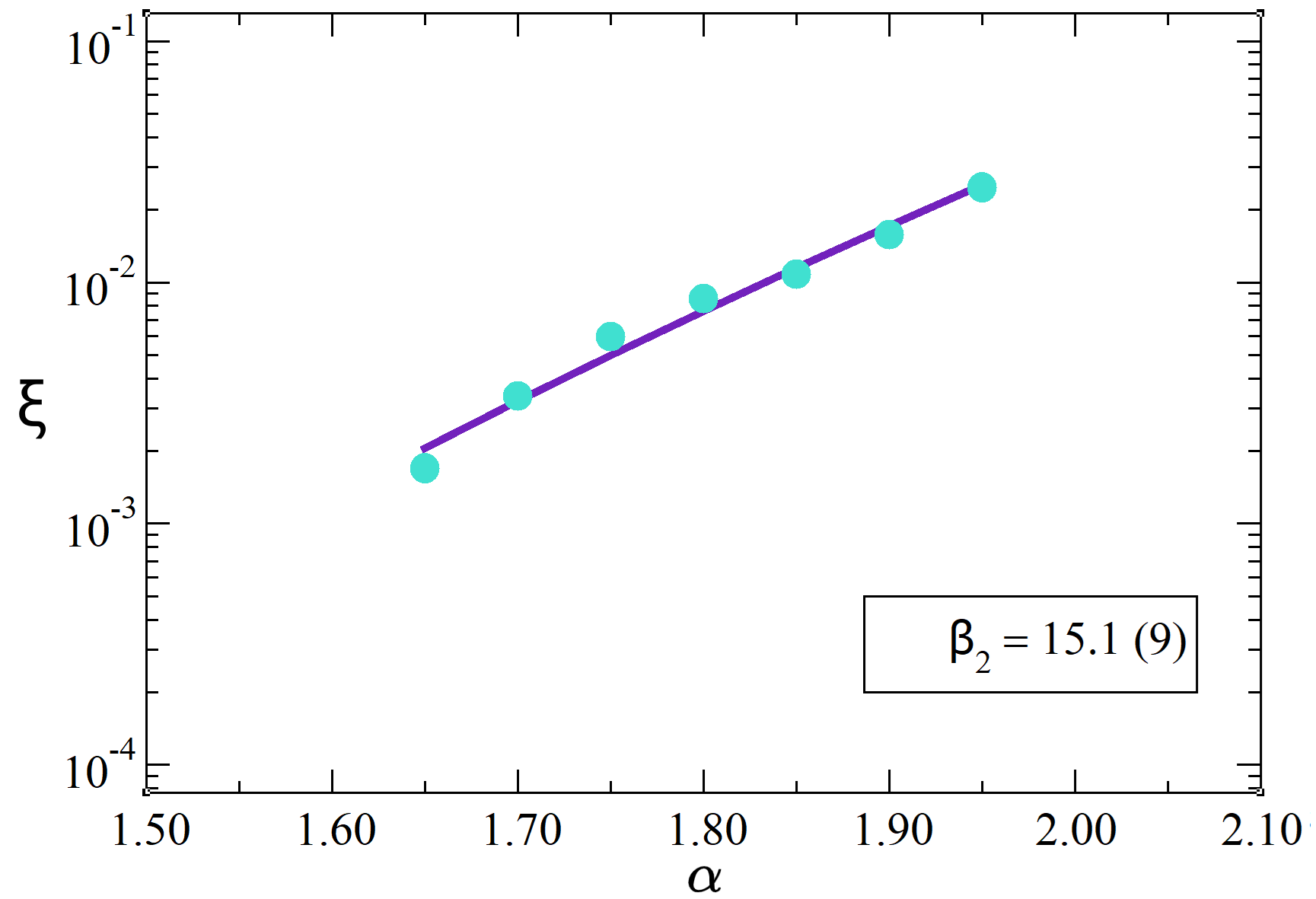} 
	\caption{The behavior of the critical exponent $\xi$ as a function of $\alpha$ is depicted. As can be seen, the relationship between $\alpha$ and the critical exponent $\xi$ is governed by a power law, such that $\xi \propto \alpha^{\beta_2}$, with an exponent $\beta_2=15.1(9)$.}
	\label{Fig:ScalingInv2-b}
\end{figure}

 Now, lets vary  $\alpha$, keeping $K=23.0602$ and $h=10$ as constants. As one can notice in Fig. \ref{Fig:ScalingInv2-a}, the shape presented for the decay of $P_S(n)$ for several values of $\alpha$ seems to be similar. This observation suggests a potential scaling invariance for this measure. To investigate this possibility, we examined $P_S(n)$ under the following scaling hypotheses:
$$\xi \propto \alpha^{\beta_2},$$
where $\beta_2$ represents a critical exponent, which can be found through the analysis of the numerical fitting of $\xi~vs.~\alpha$, as shown in Fig. \ref{Fig:ScalingInv2-b}, such that it is given by $\beta=15.1(9)$. Thus, knowing the value of $\beta$, one can rescale the horizontal axis by transforming $n \rightarrow n \alpha^{\beta_2}$. This transformation results in overlapping curves as depicted in Fig. \ref{Fig:ScalingInv2-a}, converging into a unified and universal plot in Fig. \ref{Fig:ScalingInv2-c}. This outcome signifies the scaling invariance for $P_S(n)$ concerning the parameter $\alpha$.

\par
\begin{figure*}[!ht]
	\centering
	\begin{subfigure}[ht]{0.48\textwidth}
		\centering
		\includegraphics[width=\textwidth]{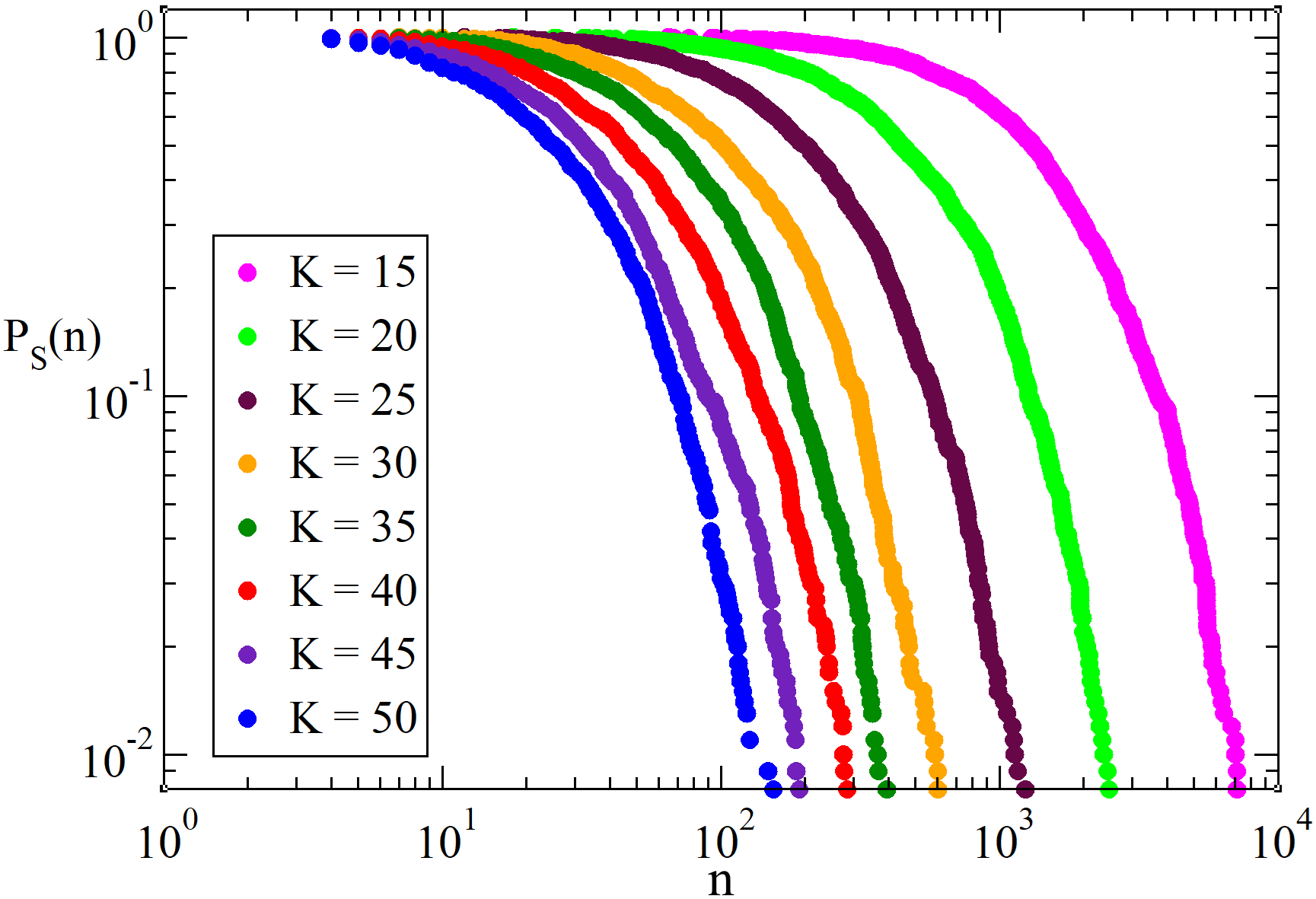}
		\caption{ }
		\label{Fig:ScalingInv3-a}
	\end{subfigure}
	\hfill
	\begin{subfigure}[ht]{0.48\textwidth}
		\centering
		\vspace{0.02cm}
		\includegraphics[width=\textwidth]{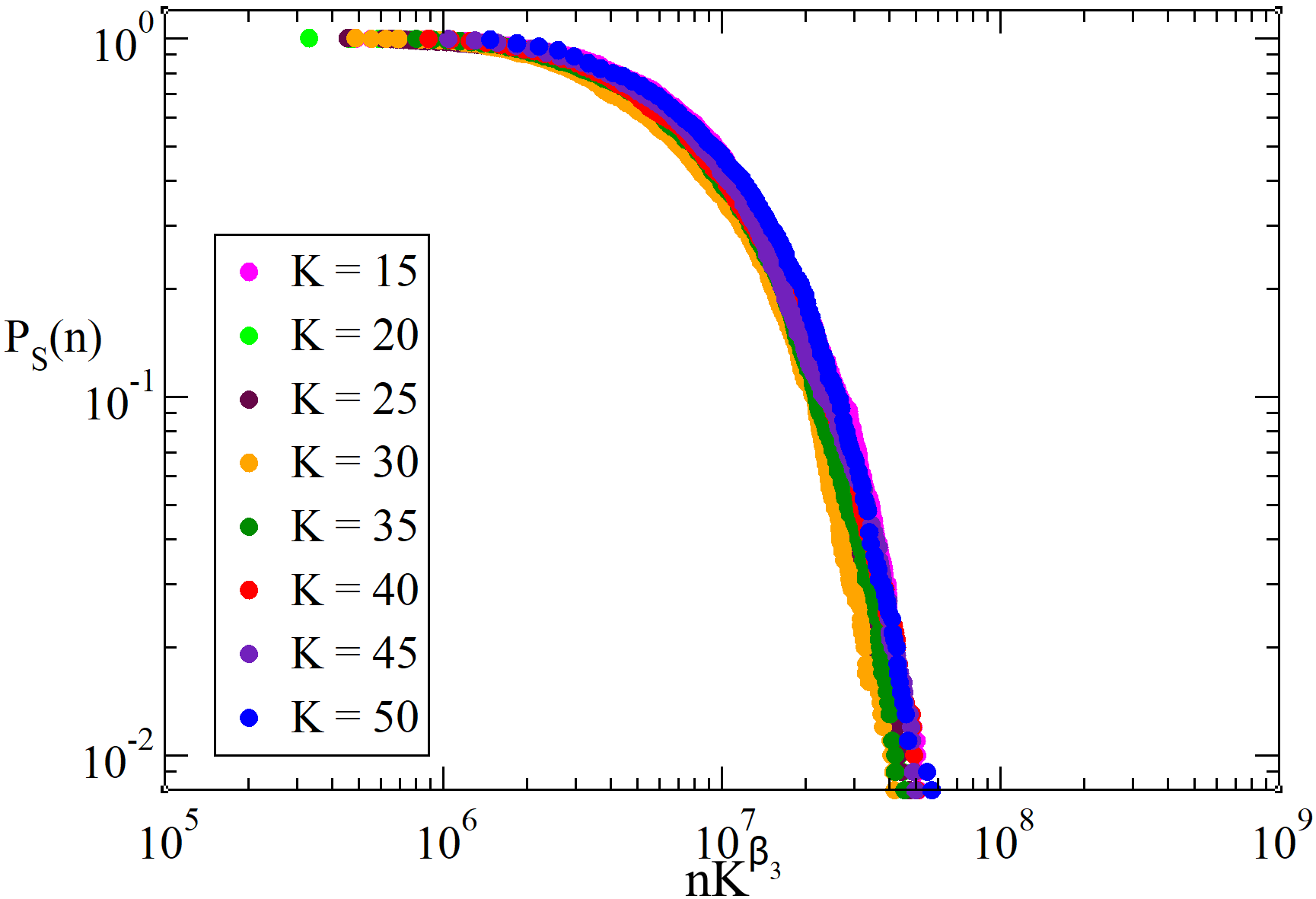}
		\caption{ }
		\label{Fig:ScalingInv3-c}
	\end{subfigure}
	\caption{(a) Plot of the survival probability curves $P_S (n)$ as functions of the number of collisions $n$, considering $K$ in range $K\in[15,50]$, $\alpha=1.7$, $h=100$ and computed for $M=10^3$ orbits. The curves can be described by an exponential $P_S (n) \propto \exp(-\xi n)$; (b) We display the overlapping of the $P_S (n)$ curves in a single universal curve, indicating the scaling invariance of the survival probability concerning variations of $K$.}
\end{figure*}

\begin{figure}[ht]
	\centering
	\includegraphics[width=0.48\textwidth]{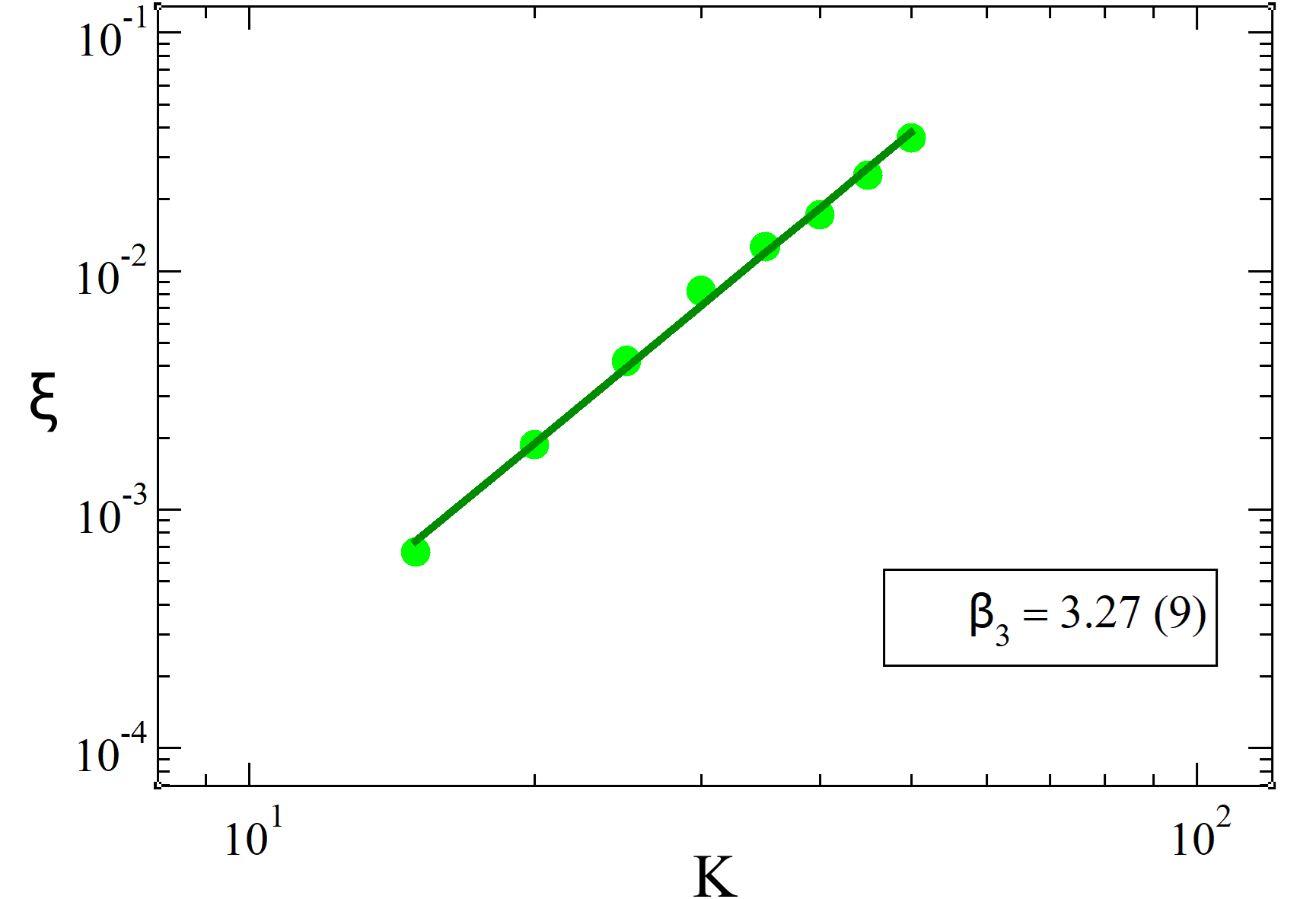} 
	\caption{The behavior of the critical exponent $\xi$ as a function of $K$ is depicted. As can be seen, the relationship between $\alpha$ and the critical exponent $\xi$ is governed by a power law, such that $\xi  \propto K^{\beta_3}$, with an exponent $\beta_3=3.27(9)$.}
	\label{Fig:ScalingInv3-b}
\end{figure}

 Finally, after the preliminary results indicated that the survivor probability is scaling invariant concerning $h$ and $\alpha$, a question is raised: \textit{Is the behavior of the survival probability $P_S(n)$ scale invariant due to variations of $K$?}. To answer this, the analysis performed in Fig. \ref{Fig:ScalingInv1-a} and Fig. \ref{Fig:ScalingInv2-b} is repeated, considering the following scaling hypotheses: 
$$\xi \propto K^{\beta_3}.$$ 
In Fig. \ref{Fig:ScalingInv3-b}, it is found that the value of the critical exponent is $\beta_3=3.27(9)$. Thus, the curves in Fig. \ref{Fig:ScalingInv3-a} overlap in a very good fashion after an appropriate scaling transformation, as shown in Fig. \ref{Fig:ScalingInv3-c}, confirming that the mapping presents scale invariance about $K$ and answering the question proposed previously.

\section{Discussion} \label{Sec:Discussion}

\hspace{0.5cm} In this paper, some dynamical and statistical properties of a fractional map were studied. Initially, a mapping was derived from a fractional differential equation by considering the Caputo derivative in the ODE that originates the Chirikov map. Thus, using an escape formalism, the behavior of orbits in the previously obtained mapping, referred to as the Caputo Fractional Standard Map (CFSM), was investigated. This map is parameterized by the control parameters $K$ and $\alpha$. In the study of survival probability, a new control parameter $h$ is considered, and this observable now depends on $(K, \alpha, h)$. Varying these parameters, $P_s(n)$ presents an initial plateau for short times, followed by an exponential decay for medium and long times, as illustrated in Fig. \ref{Fig:ScalingInv1-a}, \ref{Fig:ScalingInv2-a}, and \ref{Fig:ScalingInv3-a}. The behavior of survival probability allows us to propose that it is described by $P_S(n) \propto e^{-\gamma}$, where $\gamma$ is a critical exponent that depends on the values of the assumed control parameters, as displayed in Fig. \ref{Fig:ScalingInv1-b}, \ref{Fig:ScalingInv2-b}, and \ref{Fig:ScalingInv3-b}.

\par The overlap of the curves shown in Fig. \ref{Fig:ScalingInv1-c}, \ref{Fig:ScalingInv2-c} and \ref{Fig:ScalingInv3-c} confirms a scaling invariance observed for Survival probability when studied in the context of Caputo Fractional Standard Mapping (CFSM), described in Equation \ref{19}, for all control parameters ($\alpha$, $K$ and $h$).

\section*{Acknowledgments}
D.B. reports financial support was provided by FAPESP (No.2022/03612-6). We would like to acknowledges Edson Denis Leonel (EDL) for  valuable discussions and Luis Renato Damin (LRD) for careful reading on the text.

\end{document}